\documentstyle[preprint,aps]{revtex}
\tighten
\begin{document}
\draft
\preprint{\vbox{Submitted to Nuclear Physics A 
                      \hfill FSU-SCRI-96-08}}
\title{The Okamoto-Nolen-Schiffer anomaly without 
       \boldmath{$\!\!\rho$}-\boldmath{$\omega$} mixing}
\author{F. Cardarelli}
\address{Department of Physics and Supercomputer Computations 
         Research Institute, \\
         Florida State University, Tallahassee, FL 32306}
\author{J. Piekarewicz}
\address{Supercomputer Computations Research Institute, \\
         Florida State University, Tallahassee, FL 32306}
\date{\today}
\maketitle
 
\begin{abstract}
 We examine the effect of isospin-violating meson-nucleon 
coupling constants and of $\pi$-$\eta$ mixing on the 
binding-energy differences of mirror nuclei in a model 
that possesses no contribution from $\rho$-$\omega$ mixing.
The ${}^{3}$He-${}^{3}$H binding-energy difference is computed 
in a nonrelativistic approach using a realistic wave function.
We find the ${}^{3}$He-${}^{3}$H binding-energy difference
very sensitive to the short-distance behavior of the 
nucleon-nucleon potential. We conclude that for the typically 
hard Bonn form factors such models can not account for the 
observed binding-energy difference in the three-nucleon system. 
For the medium-mass region (A=15--41) the binding-energy 
differences of mirror nuclei are computed using a relativistic 
mean-field approximation to the Walecka model. We obtain large 
binding-energy differences---of the order of several hundred 
keV---arising from the pseudoscalar sector. Two effects are 
primarily responsible for this new finding: a) the inclusion 
of isospin breaking in the pion-nucleon coupling constant, and 
b) the in-medium enhancement of the small components of the 
bound-state wave functions. We look for off-shell ambiguities 
in these results and find them to be large.

\end{abstract}

\narrowtext
 
\section{Introduction}
\label{sec:intro}

     Some of our most fundamental ideas about charge-symmetry-breaking 
(CSB) phenomena in nuclear physics~\cite{henmil79,miller90} are currently 
being revisited. To a large extent this revision has been prompted by 
the suggestion of Goldman, Henderson, and Thomas of a suppressed 
contribution from $\rho$-$\omega$ mixing to the CSB component of the 
nucleon-nucleon (NN) interaction at small momentum transfers~\cite{ght92}. 
Since then, many calculations, using a variety of theoretical approaches, 
have confirmed this 
suppression~\cite{piewil93,krein93,hats94,mitch94,oconn94,malta95}.
Although the issue continues to be 
controversial~\cite{miller94,oconn95}, the search for alternate CSB 
mechanisms has already started. Indeed, in a recent publication we 
have proposed isospin breaking in the meson-nucleon vertices as an 
additional source of isospin violation~\cite{ghp95}. We have used 
isospin breaking in the pion-nucleon coupling constant~\cite{piekar95} 
to account for the large isospin violation in the pion-nucleon
system reported recently by Gibbs, Ai, and Kaufmann~\cite{gak95}. 
Moreover, we have studied the impact of isospin violation in the 
vector-meson--nucleon coupling constants on the difference between 
the neutron and proton analyzing powers ($\Delta A$) measured in 
elastic $\vec{n}$-$\vec{p}$ scattering~\cite{knut90}. We have 
concluded that the magnitude of the resulting class IV CSB potential 
was consistent with that phenomenologically required by the 
experiment~\cite{ghp95}. 

 However, other CSB observables, such as the scattering-length 
differences of the $pp$ and $nn$ systems~\cite{slaus89}, and
the binding-energy differences of mirror 
nuclei---the Okamoto-Nolen-Schiffer (ONS) 
anomaly~\cite{oka64,nolsch69}---are particularly insensitive 
to the class IV component of the CSB potential. Rather, these
observables are driven by the class III component. 
Indeed, several authors~\cite{bluiqb87,coon87,ishsas90} 
have observed that the class III component of the CSB 
potential---generated from ``on-shell'' $\rho$-$\omega$ 
mixing---dominates the binding-energy discrepancy of mirror 
nuclei and could explain most of the ONS anomaly~\cite{miller90}.

 Motivated by the recent developments, we propose to study 
the ONS anomaly in a model which possesses no contribution from 
$\rho$-$\omega$ mixing; in our model, the resulting CSB potential
is generated exclusively from $\pi$-$\eta$ mixing and from isospin 
violation in the meson-nucleon coupling constants. To compute the 
${}^{3}$He-${}^{3}$H binding-energy difference we employ a 
realistic three-body wave function obtained from the variational 
analysis of Ref.~\cite{pace77} using the Reid-soft core potential.
To study the ONS anomaly in the A=15--41 region we use  
a relativistic mean-field approximation to the Walecka 
model~\cite{serwal86}. Although the Walecka model has been used 
recently to compute binding-energy differences in mirror 
nuclei~\cite{nedroo88,bgk95,kkb96}, these studies have been limited 
to the vector sector: Coulomb corrections plus on-shell 
$\rho$-$\omega$ mixing. However, since these contributions are 
dominated by the timelike part of the vector vertex ($\gamma^{0}$) 
they are not particularly sensitive to relativistic effects.
In contrast, one expects important relativistic corrections in 
the pseudoscalar sector because of the structure of the
($\gamma^{5}$) vertex. Examining the role of relativistic effects 
and of isospin violation in the meson-nucleon coupling constants 
are two of the major components of the present investigation.

 Our paper has been organized as follows. In Sec.~\ref{sec:csbpot}
we introduce the CSB potentials derived from meson mixing and from 
isospin violation in the meson-nucleon coupling constants. There,
we review some of the ideas germane to our model. 
Sec.~\ref{sec:threebody} is reserved to the analysis of the ONS
anomaly in the three-nucleon system with special emphasis placed
on the sensitivity of our results to the choice of hadronic form
factors. Binding-energy differences of A=15--41 mirror nuclei are 
computed in Sec.~\ref{sec:manybody} using various off-shell 
extrapolations of the NN potential. Finally, we present our 
conclusions in Sec.~\ref{sec:concl}.

\section{Charge-symmetry-breaking Potentials}
\label{sec:csbpot} 

\subsection{Vector-meson sector}

 The CSB potentials which arise from isospin violation in the
vector-meson--nucleon vertices and from $\rho$-$\omega$ mixing 
are given, respectively, by~\cite{henmil79,miller90,ghp95}
\begin{mathletters}
 \label{vveccsb}
 \begin{eqnarray}
  \widehat{V}^{\omega}_{\rm CSB} &=& 
   {V}^{\omega}_{\rm CSB}(Q^{2}) 
   \Big[ \Gamma^{\mu}(1) \gamma_{\mu}(2) \tau_z(1) +
         \gamma^{\mu}(1) \Gamma_{\mu}(2) \tau_z(2) \Big] 
   \;, \label{vomegaa} \\
  \widehat{V}^{\rho}_{\rm CSB} &=& 
   {V}^{\rho}_{\rm CSB}(Q^{2})
   \Big[ \Gamma^{\mu}(1) \gamma_{\mu}(2) \tau_z(2) +
         \gamma^{\mu}(1) \Gamma_{\mu}(2) \tau_z(1) \Big] +   
        \label{vrhoa} \\
   &\phantom{+}&
   {V}^{\prime\rho}_{\rm CSB}(Q^{2}) 
   \left[\Gamma^{\mu}(1) \Gamma_{\mu}(2) 
         \Big(\tau_z(1)+\tau_z(2)\Big) \right] \;, \nonumber \\  
  \widehat{V}^{\rho\omega}_{\rm CSB} &=& 
   {V}^{\rho\omega}_{\rm CSB}(Q^{2}) 
   \left[\gamma^{\mu}(1) \gamma_{\mu}(2) 
         \Big(\tau_z(1)+\tau_z(2)\Big) \right] +
   \label{vrhoomegaa} \\
   &\phantom{+}&
   {V}^{\prime\rho\omega}_{\rm CSB}(Q^{2})
   \Big[\Gamma^{\mu}(1) \gamma_{\mu}(2) \tau_z(1) +
        \gamma^{\mu}(1) \Gamma_{\mu}(2) \tau_z(2)\Big] \;. \nonumber 
 \end{eqnarray}
\end{mathletters}
where $Q^{2} \equiv -q_{\mu}^{2}$ is the negative of the
four-momentum transfer, $\Gamma^{\mu} \equiv 
i\sigma^{\mu\nu}(p'-p)_{\nu}/2M$, and we have defined
\begin{mathletters}
 \begin{eqnarray}
 {V}^{\omega}_{\rm CSB}(Q^{2}) &\equiv& 
   {g_{\lower 2pt \hbox{$\scriptstyle NN\omega$}}^2 
    \over Q^{2} + m_{\omega}^2} f_{1}^{\omega}
    \;, \label{vomegab} \\
 {V}^{\rho}_{\rm CSB}(Q^{2}) &\equiv& 
   {g_{\lower 2pt \hbox{$\scriptstyle NN\rho$}}^2 
    \over Q^{2} + m_{\rho}^2} f_{0}^{\rho}
    \;, \label{vrhob} \\
 {V}^{\prime\rho}_{\rm CSB}(Q^{2}) &\equiv& 
   {g_{\lower 2pt \hbox{$\scriptstyle NN\rho$}} 
    f_{\lower 2pt \hbox{$\scriptstyle NN\rho$}} 
    \over Q^{2} + m_{\rho}^2} f_{0}^{\rho}
    \;, \label{vrhopb} \\
 {V}^{\rho\omega}_{\rm CSB}(Q^{2}) &\equiv& -
  {g_{\lower 2pt \hbox{$\scriptstyle NN\rho$}}
   g_{\lower 2pt \hbox{$\scriptstyle NN\omega$}}
   \over (Q^{2} + m_{\rho}^2)(Q^{2} + m_{\omega}^2)} 
   \langle \rho | H | \omega \rangle 
   \;,  \label{vrhoomegab} \\
 {V}^{\prime\rho\omega}_{\rm CSB}(Q^{2}) &\equiv& -
  {f_{\lower 2pt \hbox{$\scriptstyle NN\rho$}}
   g_{\lower 2pt \hbox{$\scriptstyle NN\omega$}}
   \over (Q^{2} + m_{\rho}^2)(Q^{2} + m_{\omega}^2)} 
   \langle \rho | H | \omega \rangle \;.
 \label{vrhoomegapb} 
 \end{eqnarray}
\end{mathletters}
Note that $g_{\lower 2pt \hbox{$\scriptstyle NN\omega$}}$,
$g_{\lower 2pt \hbox{$\scriptstyle NN\rho$}}$, and
$f_{\lower 2pt \hbox{$\scriptstyle NN\rho$}}$, are the 
isospin-conserving meson-nucleon coupling constants
extracted from fits to two nucleon data and displayed
in Table~\ref{tableone} using the Bonn B potential
model~\cite{machl89}. 
Charge-symmetry breaking in the potentials is 
characterized by the appearance of three isospin-violating 
parameters: $f_{1}^{\omega}$, $f_{0}^{\rho}$, and 
$\langle\rho|H|\omega\rangle$; an isovector-tensor 
$NN\omega$ coupling constant, an isoscalar-tensor 
$NN\rho$ coupling constant, and the $\rho$-$\omega$ 
mixing amplitude, respectively. 

 Our model is inspired by the notion of vector-meson dominance (VMD), 
so that the vector mesons couple to the appropriate isospin components 
of a conserved electromagnetic current. In models of this kind the 
$\rho$-$\omega$ mixing amplitude necessarily vanishes at 
$Q^{2}~=~0$~\cite{piewil93,oconn94}. Moreover, our VMD assumption 
places important constraints on the form of the isospin-violating
vector-meson--nucleon vertex at $Q^{2}=0$. Specifically, isospin 
breaking can occur only in the tensor part of the vertex; the 
vector coupling is protected by gauge 
invariance~\cite{ghp95,dmitra95,maltb95}. Thus, gauge invariance 
precludes the appearance of a class III---vector-vector---component 
in the CSB potential generated from vector-meson exchange. For the 
analysis of $\Delta A$ this fact is of no consequence, as class III 
CSB potentials do not contribute to $np$ observables. Indeed, the 
class IV contribution from $\omega$-meson exchange is able to fill 
the role demanded by data; recall that this term is identical in 
structure and comparable in size to the one generated from
on-shell $\rho$-$\omega$ mixing~\cite{ghp95}. However, 
for observables in which the vector-vector component of 
$\widehat{V}^{\rho\omega}_{\rm CSB}$ was believed to be 
dominant---such as in the case of the ONS
anomaly~\cite{bluiqb87,coon87,ishsas90}---the absence of the 
corresponding term from $\widehat{V}^{\omega}_{\rm CSB}$ and
$\widehat{V}^{\rho}_{\rm CSB}$ could have important 
phenomenological consequences. This conclusion is unavoidable: 
it follows directly from gauge invariance; gauge invariance forces 
the isospin-violating vector coupling and the $\rho$-$\omega$ mixing 
amplitude to vanish at $Q^{2}=0$. 

\subsection{Pseudoscalar-meson sector}

 The CSB potential which emerges from isospin violations in the
pseudoscalar-meson sector is given by,
\begin{mathletters}
 \label{vpscsb}
 \begin{eqnarray}
  &&
  \widehat{V}^{(5)}_{\rm CSB} = 
   {V}^{(5)}_{\rm CSB}(Q^{2}) \gamma^{5}(1)\gamma^{5}(2)
   \Big[\tau_z(1)+\tau_z(2)\Big] \;, \\
  && 
   {V}^{(5)}_{\rm CSB}(Q^{2}) =
   {V}^{\pi}_{\rm CSB}(Q^{2}) + 
   {V}^{\pi\eta}_{\rm CSB}(Q^{2}) \;,
 \end{eqnarray}
\end{mathletters}
with ${V}^{(5)}_{\rm CSB}$ made out of contributions arising
from isospin violation in the pion-nucleon vertex and from 
$\pi$-$\eta$ mixing~\cite{ghp95}, respectively, 
\begin{mathletters}
 \begin{eqnarray}
 {V}^{\pi}_{\rm CSB}(Q^{2}) &\equiv& 
   {g_{\lower 2pt \hbox{$\scriptstyle NN\pi$}}^2 
    \over Q^{2} + m_{\pi}^2} g_{0}^{\pi}
    \;, \label{vpion} \\
 {V}^{\pi\eta}_{\rm CSB}(Q^{2}) &\equiv& -
  {g_{\lower 2pt \hbox{$\scriptstyle NN\pi$}}
   g_{\lower 2pt \hbox{$\scriptstyle NN\eta$}}
   \over (Q^{2} + m_{\pi}^2)(Q^{2} + m_{\eta}^2)} 
   \langle \pi | H | \eta \rangle \;.
 \end{eqnarray}
\end{mathletters}

 The isospin-conserving pion-nucleon coupling constant 
$g_{\lower 2pt \hbox{$\scriptstyle NN\pi$}}$ is given 
by the Bonn B potential value listed in Table~\ref{tableone}.
For the $NN\eta$ coupling constant we have decided to use a
value extracted from SU(3)-flavor symmetry~\cite{coon82},
rather than the Bonn-potential value of 
$g_{\lower 2pt \hbox{$\scriptstyle NN\eta$}}^{2}/4\pi=2.25$.
Indeed, it is believed that the Bonn potential overestimates
the coupling; a recent analysis based on $\eta$-photoproduction
data suggests values as low as 
$g^2_{NN\eta}/4\pi \alt 0.5$~\cite{tiator94} (see also
Ref.~\cite{piekar93}). For the $\eta N$ cutoff parameter 
we have simply assumed: $\Lambda_{NN\eta}=\Lambda_{NN\pi}$. 
The two parameters driving the CSB potential are the 
isospin-violating pion-nucleon coupling constant ($g_{0}^{\pi}$)
and the $\pi$-$\eta$ mixing amplitude ($\langle\pi|H|\eta\rangle$).
CSB potentials derived from $\pi$-$\eta$ mixing have been included 
in previous nonrelativistic analyses of the ONS 
anomaly~\cite{bluiqb87,ishsas90,coon82}. However, to our knowledge, 
CSB potentials arising from isospin violation in the meson-nucleon 
coupling constants have never been incorporated into these analyses, 
and only recently have they been used to study isospin violations 
in low-energy pion-nucleon scattering~\cite{piekar95}. 

\section{Binding-energy difference in the A=3 system}
\label{sec:threebody} 

In this section we examine the effect of $\pi$-$\eta$ mixing and
of isospin-violating meson-nucleon coupling constants on the 
${}^{3}$He-${}^{3}$H binding-energy difference. Experimentally, 
this difference is very precisely known and has the value of
764~keV. A nearly model-independent method for estimating the 
electromagnetic contribution to the binding-energy difference
has already been developed~\cite{coon87,friar70}; ambiguities remain, 
however, on how to separate the meson-exchange-current component
from the experimental charge form factors. These estimates suggest
an electromagnetic contribution to the binding-energy of
$693 \pm 19\pm5$~keV, leaving $71\pm19\pm5$~keV to be explained
by charge-symmetry-breaking effects\cite{miller90,coon87}. Note that
a very similar result has been obtained recently by Ishikawa and 
Sasakawa using a different method~\cite{ishsas90}. These authors 
have also computed the contribution to the binding-energy
difference arising from (on-shell) $\rho$-$\omega$ mixing and 
from $\pi$-$\eta$ mixing, with the former playing the dominant 
role. Our model possesses no contribution from $\rho$-$\omega$ 
mixing. Hence, we would like to investigate the possible role 
of alternate CSB mechanisms in accounting for the non-electromagnetic 
contribution to the ${}^{3}$He-${}^{3}$H binding-energy difference:
\begin{equation}
  \Delta E = \langle {}^{3}He | V_{pp}   | {}^{3}He \rangle 
           - \langle {}^{3}H  | V_{nn}   | {}^{3}H  \rangle 
    \equiv - \langle {}^{3}He | \Delta V | {}^{3}He \rangle \;.
\end{equation}

We shall compute the ${}^{3}$He-${}^{3}$H binding-energy difference 
using the realistic three-body wave function obtained from the 
variational analysis of Numberg, Prosperi, and Pace\cite{pace77}.
These authors have written the nuclear Hamiltonian
\begin{equation}
  H=\frac{1}{2M}({\bf p}_a^2 + 3{\bf p}_b^2) + 
  \sum_{i,j=1 \atop (i<j)}^{3} V_{ij} \;,
\end{equation}
in terms of the two intrinsic coordinates 
\begin{equation}
{\bf a} = \sqrt{1 \over 2}({\bf r}_1 - {\bf r}_2) \;; \quad
{\bf b} = \sqrt{1 \over 2}(2{\bf r}_3 -{\bf r}_1 -{\bf r}_2) \;,
\end{equation}
and their corresponding conjugate momenta. Note that the 
indices 1 and 2 are reserved for the identical particles
in the system; this is particularly convenient for the 
perturbative estimate of the ${}^{3}$He-${}^{3}$H
binding-energy shift. The basic dynamical input is contained 
in the two-nucleon interaction ($V_{ij}$) which was chosen to 
be the Reid soft-core potential\cite{reid68}. Matrix elements of 
the Hamiltonian were computed using a truncated basis of 820 
harmonic-oscillator states and the ground-state wave function 
was obtained from direct matrix diagonalization. This wave 
function provides a satisfactory description of the charge form 
factors of ${}^{3}$He and ${}^{3}$H in the low-momentum-transfer 
region~\cite{pace77,ciofi79}. Moreover, a perturbative estimate of
the Coulomb shift in ${}^{3}$He yields 672~keV---a value 
consistent with that reported in Refs.~\cite{coon87,ishsas90}.

For simplicity, the binding energy difference will be evaluated 
with the dominant ${}^{1}S_{0}$ component of the trinucleon 
wave function. Note that for this case it is sufficient to evaluate
the ${}^{1}S_{0}$ component of the CSB potential:
\begin{equation}
 \Delta V \equiv V_{nn}({}^{1}S_{0}) - V_{pp}({}^{1}S_{0}) \;.
\end{equation}
The singlet CSB component of the NN potential is obtained by 
performing a nonrelativistic reduction of the various Lorentz 
structures given in Eqs.~(\ref{vveccsb},\ref{vpscsb}) and by 
evaluating the meson propagators in the static 
limit~\cite{henmil79,coon75}. We obtain ($f_1^{\rho}\equiv
f_{\lower 2pt \hbox{$\scriptstyle NN\rho$}}/
g_{\lower 2pt \hbox{$\scriptstyle NN\rho$}}$)
\begin{mathletters}
 \begin{eqnarray}
  \Delta V^{\pi\eta}({\bf q}) &=&  
   g_{\lower 2pt \hbox{$\scriptstyle NN\pi$}}
   g_{\lower 2pt \hbox{$\scriptstyle NN\eta$}}
  \langle\pi|H|\eta\rangle {{\bf q}^{2}/M^{2} \over 
  {({\bf q}^2+m_{\pi}^2)}{({\bf q}^2+m_{\eta}^2})}          \;, \\
  \Delta V^{\pi}({\bf q}) &=& - 
   g_{\lower 2pt \hbox{$\scriptstyle NN\pi$}}^{2}g_0^{\pi}
     {{\bf q}^{2}/M^{2} \over {{\bf q}^2+m_{\pi}^2}}        \;, \\
  \Delta V^{\rho}({\bf q}) &=& - 
   g_{\lower 2pt \hbox{$\scriptstyle NN\rho$}}^{2}
  (1+2f_1^{\rho}) f_0^{\rho}
     {{\bf q}^{2}/M^{2} \over {{\bf q}^2+m_{\rho}^2}}       \;, \\
  \Delta V^{\omega}({\bf q}) &=& - 
   g_{\lower 2pt \hbox{$\scriptstyle NN\omega$}}^{2}f_1^{\omega}
     {{\bf q}^{2}/M^{2} \over {{\bf q}^2+m_{\omega}^2}}         \;. 
 \end{eqnarray}
\end{mathletters}
Note the appearance of the ``relativistic-correction'' factor 
${\bf q}^{2}/M^{2}$. This factor emerges from the allowed 
Lorentz structures present in the CSB potentials. Recall 
that---in contrast to previous analyses based on on-shell 
$\rho$-$\omega$ mixing---no Lorentz vector structure 
[$\gamma^{\mu}(1)\gamma_{\mu}(2)$] is allowed in our model.  
For the $\pi$-$\eta$ mixing amplitude we have used the accepted 
value of $\langle\pi|H|\eta\rangle=-4200$~MeV$^{2}$~\cite{coon82}. 
For the isospin-violating meson-nucleon parameter we have adopted 
the nonrelativistic quark-model estimate of Refs.~\cite{ghp95}:
\begin{mathletters}
 \begin{eqnarray}
  g_{0}^{\pi}&=&{3 \over 10}{\Delta m \over m}   \approx 0.004 \;, \\
  f_{0}^{\rho}&=&{3 \over 2}{\Delta m \over m}   \approx 0.020 \;, \\
  f_{1}^{\omega}&=&{5 \over 6}{\Delta m \over m} \approx 0.011 \;,
 \end{eqnarray}
\end{mathletters} 
where $m=313$~MeV is the average constituent quark mass and 
$\Delta m=4.1$~MeV is the down-up quark mass difference~\cite{licht89}. 
Although these values are model dependent, they are insensitive to the 
spatial component of the nucleon wave function; they follow directly 
from its spin and flavor content. Moreover, the value of $g_{0}^{\pi}$
is compatible with other estimates available in the 
literature~\cite{miller90} and, in particular, with the recent value 
reported by Henley and Meissner from QCD sum rules~\cite{henmei96}. 

To facilitate the interpretation of our results it is convenient 
to transform the above expressions into configuration space: 
\begin{mathletters}
 \begin{eqnarray}
  \Delta V^{\pi\eta}(r) &=&  
   {g_{\lower 2pt \hbox{$\scriptstyle NN\pi$}}
    g_{\lower 2pt \hbox{$\scriptstyle NN\eta$}} \over 4\pi}
   {\langle\pi|H|\eta\rangle \over m_{\eta}^2-m_{\pi}^2}
   \left[ 
     \left({m_{\eta} \over M}\right)^{2}{e^{-m_{\eta}r} \over r} - 
     \left({m_{\pi}  \over M}\right)^{2}{e^{-m_{\pi}r } \over r}
   \right] \;, \\
  \Delta V^{\pi}(r) &=&  
   {g_{\lower 2pt \hbox{$\scriptstyle NN\pi$}}^{2}\over 4\pi}
    g_0^{\pi}
   \left[
     \left({m_{\pi}  \over M}\right)^{2}{e^{-m_{\pi}r}  \over r} -
     {\delta(r) \over M^{2}r^{2}} 
   \right] \;, \\
  \Delta V^{\rho}(r) &=&  
   {g_{\lower 2pt \hbox{$\scriptstyle NN\rho$}}^{2}\over 4\pi}
    (1+2f_1^{\rho})f_0^{\rho}
   \left[
     \left({m_{\rho}  \over M}\right)^{2}{e^{-m_{\rho}r} \over r} -
      {\delta(r) \over M^{2}r^{2}} 
   \right] \;, \\
  \Delta V^{\omega}(r) &=&  
   {g_{\lower 2pt \hbox{$\scriptstyle NN\omega$}}^{2}\over 4\pi}
    f_1^{\omega}
   \left[
     \left({m_{\omega}  \over M}\right)^{2}{e^{-m_{\omega}r} \over r} -
     {\delta(r) \over M^{2}r^{2}} 
   \right] \;.
 \end{eqnarray}
\end{mathletters}
Note that in configuration space the ${}^{3}$He-${}^{3}$H 
binding-energy difference emerges from a simple overlap between the 
singlet component of the wave function [$R(a)$] and the CSB potential
\begin{equation}
  \Delta E = - \int_{0}^{\infty} a^2 da |R(a)|^{2} 
               \Delta V(\sqrt{2}a) \equiv 
               \int_{0}^{\infty} I(a)da \;.
 \label{iofa}
\end{equation}
We will refer to these estimates as the point-coupling results,
as hadronic form factor have yet to be introduced. We note, 
as a consequence of the relativistic factor ${\bf q}^{2}/M^{2}$, 
the appearance of a contact term [$\delta(r)$] in all single-meson 
exchanges (the contact term is absent from the $\pi$-$\eta$ mixing 
potential because of the additional meson propagator). For realistic 
wave functions---such as the one used here---the contact term does
not contribute to the overlap as a result of the repulsive short-range 
correlations. Thus, we conclude that all single-meson-exchange potentials 
generate a repulsive $\Delta V$, namely, a CSB potential that is 
attractive(repulsive) in the $pp(nn)$ channel---in contrast to the 
experimental observation. Note that this conclusion depends 
exclusively on the sign of the isospin-violating meson-nucleon 
coupling constants; recall that in our model these constants are all 
positive as a result of the up quark being lighter than the down quark. 

This conclusion could change dramatically upon the inclusion of
hadronic form factors. With form factors the contact term gets 
smeared over a distance of the order of $\Lambda^{-1}$ (with $\Lambda$ 
being the cutoff parameter) and could contribute to the overlap if this 
distance becomes larger than the size of the repulsive core. Most of 
our knowledge about hadronic form factor comes from phenomenological 
fits to a large body of two-nucleon data~\cite{machl89}. These fits 
reveal hadronic form factors that are typically harder 
($\Lambda\!\agt\!1.5$~GeV) than the corresponding ones extracted 
from electron-nucleon scattering experiments
($\Lambda\!\simeq\!0.8$~GeV). However, the subject of hadronic 
form factors is highly controversial. Indeed, (model-dependent) analyses 
of deep-inelastic-scattering (DIS) experiments suggest that the
$NN\pi$ form factor is considerably softer than 
$\Lambda_{NN\pi}\!\sim\!1.5$~GeV~\cite{fms89,kum91,sst91}. 
Moreover, Holinde and Thomas have shown recently that it is possible 
to obtain an adequate description of two-nucleon data with the softer 
pion-nucleon form factor ($\Lambda_{NN\pi}\!\simeq\!0.8$~GeV) suggested 
by the DIS analyses~\cite{holtho90}. These authors have also shown that 
the $NN\rho$ form factor can also be made softer---but no softer than 
$\Lambda_{NN\rho}\!\simeq\!1.2$~GeV---without compromising the fit. 
Hence, we find it instructive to do the calculation of the 
${}^{3}$He-${}^{3}$H binding-energy shift using, both, the hard
form factors of Table~\ref{tableone} as well as the softer ones
suggested by the analyses of Refs.~\cite{fms89,kum91,sst91,holtho90}. 
In the latter case we fix, for simplicity, both pseudoscalar cutoff 
parameters at $\Lambda_{NN\pi(\eta)}=0.8$~GeV and both vector cutoff 
parameters at $\Lambda_{NN\rho(\omega)}=1.3$~GeV. We incorporate hadronic 
form factors into the calculation by modifying the point meson-nucleon 
coupling in the following way~\cite{machl89}:
\begin{mathletters}
 \begin{eqnarray}
  g_{\lower 2pt \hbox{$\scriptstyle NN\pi(\eta)$}} \rightarrow
  g_{\lower 2pt \hbox{$\scriptstyle NN\pi(\eta)$}}({\bf q}^{2})&=&
  g_{\lower 2pt \hbox{$\scriptstyle NN\pi(\eta)$}}
 (1+{\bf q}^2/\Lambda_{\pi(\eta)}^2)^{-1} \;, \\
  g_{\lower 2pt \hbox{$\scriptstyle NN\rho(\omega)$}} \rightarrow
  g_{\lower 2pt \hbox{$\scriptstyle NN\rho(\omega)$}}({\bf q}^{2})&=&
  g_{\lower 2pt \hbox{$\scriptstyle NN\rho(\omega)$}}
 (1+{\bf q}^2/\Lambda_{\rho(\omega)}^2)^{-2} \;.
 \end{eqnarray}
\end{mathletters}

In Table~\ref{tabletwo} we report the contribution to the 
${}^{3}$He-${}^{3}$H binding-energy difference arising
from $\pi$-$\eta$ mixing, $\pi-$, $\rho-$, and $\omega-$meson
exchange, respectively. For comparison, we have also included 
the contribution arising from $\rho$-$\omega$ mixing, with the
mixing amplitude fixed at its on-shell value:
$\langle\rho|H|\omega\rangle=-4520$~MeV$^{2}$~\cite{coon87}. 
The first column of numbers displays the 
results in the point-coupling limit. As suggested, all single 
meson-exchange potentials give rise to a negative binding-energy 
shift; only $\pi$-$\eta$ mixing leads to a positive---albeit very 
small---contribution. The binding energy-shifts computed with
the Bonn B hadronic form factors are reported in Table~\ref{tabletwo} 
under the heading of hard form factors. Evidently, the binding-energy
shift is very sensitive to the short-distance behavior of the
potential; the inclusion of form factors results in a 40\% 
reduction relative to the point-coupling estimate. This reduction
is best appreciated by plotting the integrand from which $\Delta E$
is obtained, that is $I(a)$ in Eq.~(\ref{iofa}). In Fig.~\ref{figone}
we display the contribution from one-pion exchange to $I(a)$; its
contribution to $\Delta E$ for each estimate, which is the area
under the curve, appears in parentheses next to its label. The
solid line represents the point-coupling result. In this case the
contact term does not contribute and the integrand is 
negative-definite over the whole region. As the (hard) Bonn form 
factor is introduced (dashed line) the smeared contact term dominates 
the short-distance region and cancels some of the negative contribution 
arising from the long-range term. It is then clear that for a 
sufficiently soft form factor the node in the potential---arising
from the competition between the short- and long-range pieces---could 
be moved to a large enough distance as to make the integral positive;
this is precisely what is observed when the softer form factor is
introduced (dot-dashed line). The complete set of numbers with soft 
form factors has been collected on Table~\ref{tabletwo}. For completeness 
we have also displayed in Fig.~\ref{figtwo} the corresponding integrand 
for the omega meson. The same features are evident. 
In the present case, however, the integral is negative in all three 
cases because the vector form factor remains fairly hard. Nevertheless, 
we should mention that even for the case of (perhaps) unrealistically 
soft form factors, namely $\Lambda_{NN\pi(\eta)}=0.6$~GeV and 
$\Lambda_{NN\rho(\omega)}=0.85$~GeV, our model can only account 
for +40 keV of the ${}^{3}$He-${}^{3}$H binding-energy difference; 
a result that is still one standard deviation away from the 
experimental value.

 In conclusion, we have shown that an analysis based on $\pi$-$\eta$
mixing and on isospin-violating meson-nucleon coupling constants---but
one without $\rho$-$\omega$ mixing---can not describe the 
${}^{3}$He-${}^{3}$H binding-energy difference. However, we have 
established that this finding is extremely sensitive to the 
least understood feature of the NN potential, namely, its short-distance 
structure. Thus, it is difficult to draw any definite conclusion until 
an accurate determination of the cutoff parameters in the meson-nucleon 
form factors is made---a very difficult task indeed.

\section{Binding-energy differences in A=15--41 mirror nuclei}
\label{sec:manybody} 

 In this section we examine the effect of meson mixing and of 
isospin violation in the meson-nucleon coupling constants on
the binding-energy differences of medium-mass mirror nuclei
using a relativistic mean-field approximation to the Walecka 
model~\cite{serwal86}. The Walecka model is a renormalizable 
strong-coupling field theory of nucleons interacting via the 
exchange of scalar and vector mesons. One of the great virtues 
of the model is its simplicity. Indeed, already at the mean-field 
level the Walecka model rivals some of the best available 
nonrelativistic calculations and provides a unified description 
of various nuclear properties, such as nuclear saturation, the 
spin-orbit force, and the density dependence of the nuclear 
interaction. The mean-field approximation is characterized by 
the existence of large Lorentz scalar and vector components that 
are responsible for a substantial in-medium enhancement of the 
small (lower) components of the single-particle wave functions; 
the so-called $M^{\star}$ effect. 

\subsection{Vector-meson sector}

 In our model $\rho$-$\omega$ mixing can not contribute to the 
ONS anomaly, at least at $Q^{2}=0$. Yet, a calculation of 
binding-energy shifts using the on-shell value for the 
$\rho$-$\omega$ mixing amplitude is still useful, as it provides 
a baseline against which alternate CSB mechanisms can be tested. 
Thus, we start this section by computing the impact of the dominant 
class III vector-vector component of 
$\widehat{V}^{\rho\omega}_{\rm CSB}$ to the ONS anomaly.
The purpose of this exercise is to establish the size of the 
phenomenological gap that will have to be filled by the alternate 
CSB mechanisms. Note that we will not report the contribution to
the binding-energy differences arising from the tensor-vector
and tensor-tensor terms in Eq.~\ref{vveccsb}; for spin-saturated 
nuclei their numerical impact is small (typically less than 10\%). 

 We compute the single-particle spectrum in a relativistic 
mean-field approximation to the Walecka model~\cite{serwal86} and 
incorporate CSB corrections in the Hartree-Fock approximation, i.e., 
\begin{equation}
   \Delta E_{\alpha}^{(0)} \equiv \Delta E_{\alpha}^{H} +
                                  \Delta E_{\alpha}^{F} \;,
 \label{deltaerw}
\end{equation}
where
\begin{mathletters}
 \begin{eqnarray}
   \Delta E_{\alpha}^{H} &=& +4\sum_{\beta}^{\rm occ}
    \int {d{\bf q} \over (2\pi)^{3}}
    V_{\rm CSB}^{\rho\omega}({\bf q}) \;
    \rho_{\alpha\alpha}^{(0)\star}({\bf q})  
    \rho_{\beta\beta}^{(0)}({\bf q}) \;, \\
   \Delta E_{\alpha}^{F} &=& -4\sum_{\beta}^{\rm occ}
    \int {d{\bf q} \over (2\pi)^{3}}
    V_{\rm CSB}^{\rho\omega}({\bf q}) \;
    \Big| \rho_{\beta\alpha}^{(0)}({\bf q}) \Big|^{2} \;.
 \end{eqnarray}
 \label{deltaevec}
\end{mathletters}
Note that the factor of four appearing in the above expressions
emerges from the characteristic class III isospin factor 
$\Big[\tau_z(1)+\tau_z(2)\Big]$. In Eq.~(\ref{deltaevec}) the 
sum runs over occupied orbitals in the Fermi sea, and we have 
introduced the Fourier transform of the off-diagonal 
timelike-vector density
\begin{equation}
   \rho_{\beta\alpha}^{(0)}({\bf q}) =
    \int d{\bf x} \; 
    \overline{{\cal U}}_{\beta}({\bf x})
    \gamma^{0} e^{-i{\bf q}\cdot{\bf x}} \; 
    {\cal U}_{\alpha}({\bf x}) \;. 
 \label{rhovecba}   
\end{equation}
Several remarks are in order. First, we have ignored the 
contribution from the Dirac sea (negative-energy states) 
in the evaluation of the single-particle spectrum as well 
as in the CSB corrections to the energies. Second, for 
spherically symmetric nuclei only the timelike part 
[$\gamma^{0}(1)\gamma^{0}(2)$] of the potential contributes 
to the Hartree term; for static sources and fields, current 
conservation prohibits the appearance of a three-vector 
current~\cite{serwal86}. Thus, as far as the Hartree term 
is concerned, the CSB potential generated from 
$\rho$-$\omega$ mixing behaves as a short-range Coulomb 
potential. Third, even though the spacelike-part 
[\mbox{\boldmath$\gamma$(1)}$\cdot$\mbox{\boldmath$\gamma$(2)}]
of the potential contributes to the Fock term, it is of little 
numerical significance and has been neglected. Finally, we suppress 
vector-meson production by ignoring retardation effects in the 
meson propagators.

 For spherically symmetric nuclei the calculation of the
binding-energy shifts simplifies considerably. In this limit
the eigenstates of the Dirac equation can be classified according 
to a generalized angular momentum $\kappa$ and can be written in 
a two component representation; i.e., 
\begin{equation}
  {\cal U}_{\alpha}({\bf x}) \equiv
  {\cal U}_{n\kappa m}({\bf x})= {1 \over x}
   \left[
    \begin{array}{c}
      \phantom{i}
       g_{a}(x)
      {\cal{Y}}_{+\kappa m}(\hat{\bf{x}})    \\
       i
       f_{a}(x)
      {\cal{Y}}_{-\kappa m}(\hat{\bf{x}})    
     \end{array}
   \right] \;; \quad \Big(\alpha \equiv (a;m) = (n,\kappa;m)\Big) \;.
 \label{dirwfx}
\end{equation}
The upper and lower components are expressed in terms of 
spin-spherical harmonics defined by
\begin{equation}
  {\cal{Y}}_{\kappa m}(\hat{\bf{x}}) \equiv
  \langle\hat{\bf{x}}|l{\scriptstyle{1\over 2}}jm\rangle \;; \quad
  j=|\kappa|-{1\over 2} \;; \quad
  l=\cases{   \kappa\;,   & if $\kappa > 0$; \cr
           -1-\kappa\;,   & if $\kappa < 0$. \cr}
 \label{curlyy}
\end{equation}
The binding-energy shifts can now be easily evaluated. That is,
\begin{mathletters}
 \begin{eqnarray}
   \Delta E_{\alpha}^{H} &=& + {2 \over \pi^{2}}
   \int q^{2} dq V_{\rm CSB}^{\rho\omega}(q) 
   {\cal F}_{aH}^{(0)}(q) \;, \\
   \Delta E_{\alpha}^{F} &=& - {2 \over \pi^{2}}
   \int q^{2} dq V_{\rm CSB}^{\rho\omega}(q) 
   {\cal F}_{aF}^{(0)}(q) \;, 
 \label{dehf}
 \end{eqnarray}
\end{mathletters}
where $q \equiv |{\bf q}|$ and we have defined the following
quantities
\begin{mathletters}
 \begin{eqnarray}
  {\cal F}_{aH}^{(0)}(q) &\equiv& \sum_{b}^{\rm occ}
  (2j_{b}+1) \rho_{bb0}^{(0)}(q) \rho_{aa0}^{(0)}(q) \;, \\
  {\cal F}_{aF}^{(0)}(q) &\equiv& \sum_{b;L}^{\rm occ}
  (2j_{b}+1) \Big[\langle j_{a},-1/2;j_{b},+1/2|L,0\rangle
  \rho_{abL}^{(0)}(q)\Big]^{2} \;.
 \label{curlyf}
 \end{eqnarray}
\end{mathletters}
Note that the Fourier transform of the off-diagonal 
timelike-vector density has become
\begin{equation}
  \rho_{abL}^{(0)}(q) = \int_{0}^{\infty} dx 
  \Big[g_{a}(x)g_{b}(x)+f_{a}(x)f_{b}(x)\Big]j_{L}(qx) \;;
  \quad (l_{a}+l_{b}+L={\rm even}) \;.
  \label{rhoablvec}
\end{equation}
Using the Bonn B potential parameters of 
Table~\ref{tableone}~\cite{machl89} we obtained the 
binding-energy differences reported in 
Table~\ref{tablethree}. These values are in qualitative 
agreement with those reported by Blunden and 
Iqbal~\cite{bluiqb87} and, more recently, by Barreiro, 
Gale\~ao, and Krein~\cite{bgk95}. It is interesting to note 
that, in contrast to the long-range Coulomb potential, the 
Fock term generates a substantial exchange ``correction'' 
(of the order of 30-40\%) to the direct Hartree 
contribution. Table~\ref{tablethree} also includes the
remaining discrepancy between experiment and three theoretical 
calculations of the Coulomb displacement energy~\cite{kkb96,sato76}. 
Note that the reported nonrelativistic discrepancies 
($\Delta_{\rm DME}$ and $\Delta_{\rm SKII}$) 
differ---in some cases by a large fraction---from the recent 
relativistic estimates ($\Delta_{\rm REL}$) of Koepf, Krein, and 
Barreiro~\cite{kkb96}.

\subsection{Pseudoscalar-meson sector: pseudoscalar $NN\pi$ coupling}

 We shall now compute the binding-energy shifts that arise
from the CSB potential given in Eq.~(\ref{vpscsb}) using a
mean-field approximation to the Walecka model. Because of 
the assumed pseudoscalar vertex, parity precludes a Hartree 
correction to the energy~\cite{serwal86}. Thus, we only need 
to evaluate the Fock term. That is,  
\begin{equation}
   \Delta E_{\alpha}^{({\rm PS})} = 
   +4\sum_{\beta}^{\rm occ}
    \int {d{\bf q} \over (2\pi)^{3}}V_{\rm CSB}^{(5)}({\bf q}) \;
    \Big| \rho_{\beta\alpha}^{({\rm PS})}({\bf q}) \Big|^{2} \;;
  \label{deltaeps}
\end{equation}
where we have introduced the Fourier transform of the 
off-diagonal pseudoscalar density
\begin{equation}
   \rho_{\beta\alpha}^{({\rm PS})}({\bf q}) =
    \int d{\bf x} \; 
    \overline{{\cal U}}_{\beta}({\bf x})
    \gamma^{5} e^{-i{\bf q}\cdot{\bf x}} \; 
    {\cal U}_{\alpha}({\bf x}) \;. 
 \label{rhopsba}   
\end{equation}
 As in the previous section, the calculation simplifies 
considerably in the limit of spherically symmetric nuclei.
In this limit, we obtain a binding-energy shift given by,
\begin{equation}
   \Delta E_{\alpha}^{({\rm PS})} = + {2 \over \pi^{2}}
    \int q^{2} dq V_{\rm CSB}^{(5)}(q)
    {\cal F}_{a}^{({\rm PS})}(q) \;,
  \label{deps}
\end{equation}
where
\begin{equation}
  {\cal F}_{a}^{({\rm PS})}(q) \equiv \sum_{b;L}^{\rm occ}
  (2j_{b}+1) \Big[\langle j_{a},-1/2;j_{b},+1/2|L,0\rangle
  \rho_{abL}^{({\rm PS})}(q)\Big]^{2} \;, 
  \label{curlyfps}
\end{equation}
and with
\begin{equation}
  \rho_{abL}^{({\rm PS})}(q) = \int_{0}^{\infty} dx 
  \Big[g_{a}(x)f_{b}(x)+f_{a}(x)g_{b}(x)\Big]j_{L}(qx) \;;
  \quad (l_{a}+l_{b}+L={\rm odd}) \;.
  \label{rhoablps}
\end{equation}
 Note that the pseudoscalar density is, as expected, linear 
in the small components of the wave functions and, thus, 
highly sensitive to their in-medium enhancement. Moreover, 
the induced energy shifts are positive definite---as 
suggested by the ONS anomaly---for all single-particle states. 

 The pseudoscalar contribution to the binding-energy difference 
of mirror nuclei is displayed in Table~\ref{tablethree}. We have 
listed separately the contributions arising from isospin violation 
in the pion-nucleon coupling constant and from $\pi$-$\eta$ 
mixing---with the former accounting for about 60\% of the total. 
Note that relativistic effects induce an additional enhancement 
in the binding-energy differences, relative to the nonrelativistic 
estimates of Blunden and Iqbal~\cite{bluiqb87}. From 
Table~\ref{tablethree} one observes that the binding-energy 
differences from the pseudoscalar sector are comparable---if not 
larger---than those computed originally with $\rho$-$\omega$ mixing.
Indeed, the pseudoscalar sector---alone---accounts for about 70-85\% 
of the discrepancy between the nonrelativistic theory and experiment.

To gauge the importance of relativity we have also computed the
binding-energy differences by assuming that the lower component
of the wave function is determined from the free-space relation;
the upper component remained unchanged, apart from a small
normalization correction. These ``nonrelativistic'' values appear 
in parenthesis in Table~\ref{tablefour} next to the uncorrected
numbers. The relativistic corrections are, indeed, very significant. 
This is in sharp contrast to the relativistic enhancements---of
less than 5\%---observed for $\rho$-$\omega$ mixing; recall that 
only the dominant vector-vector component has been included, 
which is relatively insensitive to the enhancement of the small 
components of the wave function. Perhaps the result that best 
captures the essence of the present analysis is the 
${}^{39}$Ca-${}^{39}$K binding-energy difference (the 
$1d_{3/2}^{-1}$ state in Table~\ref{tablefour}). Our baseline 
value for this difference is 65~keV; this represents our best 
attempt at reproducing the result by Blunden and 
Iqbal~\cite{bluiqb87}, namely, a nonrelativistic calculation with 
$g_{0}^{\pi}\equiv 0$. However, introducing, both, relativistic 
corrections and isospin violation in the pion-nucleon coupling 
constant increases the baseline value to 435~keV; this represents 
an enhancement of more than a factor of six. 

\subsection{Pseudoscalar-meson sector: pseudovector $NN\pi$ coupling}

 The above results suggest a large contribution from the pseudoscalar 
sector to the ONS anomaly. Yet, a definite statement awaits, as these 
results could be sensitive to off-shell extrapolations. Off-shell 
ambiguities arise from the fact that different Lorentz structures for 
the $NN$--pseudoscalar-meson vertex---which are all equivalent 
on-shell---can generate vastly different predictions once the 
nucleon is off its mass shell. Indeed, a pseudoscalar ($\gamma^{5}$) 
vertex is very sensitive to the in-medium enhancement of the lower 
components of the nucleon wave function, while a pseudovector coupling 
[$\gamma^{5}\gamma^{\mu}(p'-p)_{\mu}/2M$] is not. In a model in which 
chiral symmetry is not explicitly realized---such as the one employed 
here---a pseudovector $NN\pi$ coupling seems to be preferred, as it is 
known to incorporate the correct low-energy pion-nucleon dynamics without 
sensitive cancellations~\cite{matser82}. In contrast, a recent analysis 
of eta photoproduction data from the nucleon seems to favor a pseudoscalar 
$NN\eta$ coupling~\cite{tiator94}. 

 In this section we adopt a pseudovector representation for both 
($NN\pi$ and $NN\eta$) vertices. The advantage of such a choice 
is that it provides us with a lower bound for the binding-energy 
differences of mirror nuclei within the pseudoscalar sector. This, 
combined with the upper bound set by the pseudoscalar representation,
should adequately reflect the sensitivity of our results to the various 
off-shell extrapolations. 

 For the pseudovector case the angular-momentum algebra becomes 
slightly more complicated than before due to the explicit spin 
dependence of the operator. Recall that in the pseudoscalar case 
the spin dependence is implicit through the upper-to-lower 
coupling induced by the pseudoscalar vertex. The binding-energy 
shifts computed with a pseudovector $NN\pi$ coupling are now given by
\begin{equation}
   \Delta E_{\alpha}^{({\rm PV})} = + {2 \over \pi^{2}}
    \int q^{2} dq \left({q^2 \over 4M^{2}}\right) 
    V_{\rm CSB}^{(5)}(q) {\cal F}_{a}^{({\rm PV})}(q) \;,
  \label{depv}
\end{equation}
where
\begin{equation}
  {\cal F}_{a}^{({\rm PV})}(q) \equiv \sum_{b;J}^{\rm occ}
  (2j_{b}+1) \Big[\langle j_{a},-1/2;j_{b},+1/2|J,0\rangle
  \rho_{abJ}^{({\rm PV})}(q)\Big]^{2} \;. 
  \label{curlyfpv}
\end{equation}
 The pseudovector density has been computed in the static 
[$(p'-p)^{0} \rightarrow 0$] limit and is given by 
\begin{eqnarray}
  \rho_{abJ}^{({\rm PV})}(q) = \sum_{L=J \pm 1} 
  \Big[
   &&C_{JL}(+\kappa_{a},+\kappa_{b}) 
     \int_{0}^{\infty} dx g_{a}(x)g_{b}(x)j_{L}(qx) + \nonumber \\
   &&C_{JL}(-\kappa_{a},-\kappa_{b}) 
  \int_{0}^{\infty} dx f_{a}(x)f_{b}(x)j_{L}(qx) \Big] \;;
  \quad (l_{a}+l_{b}+L={\rm even}) \;,
  \label{rhoabjpv}
\end{eqnarray}
where 
\begin{equation}
  C_{JL}(\kappa_{a},\kappa_{b}) \equiv 
  {(-1)^{(L+J+1)/2} \over (2J+1)}
  \cases{L+(l_{a}-j_{a})(2j_{a}+1)+(l_{b}-j_{b})(2j_{b}+1)\;,
         & if $L=J+1 \;;$ \cr
         J-(l_{a}-j_{a})(2j_{a}+1)-(l_{b}-j_{b})(2j_{b}+1)\;,
         & if $L=J-1 \;.$ \cr}
 \label{ccoeff}
\end{equation}
Note that in the static limit a pseudovector vertex does not 
induce an off-diagonal coupling. Hence, relativistic effects,
which appear as $(f/g)^{2}$ corrections, are negligible. Moreover, 
the extra factor of $q^{2}$ that results from the pseudovector 
coupling is expressed in units of the free nucleon mass $M$---not 
in units of $M^{\star}$ (see Eq.~\ref{depv}). Thus, the binding-energy 
differences computed with a pseudovector vertex could be reproduced 
with a pseudoscalar vertex---only in the case in which the lower 
component of the wave function be determined from the free-space 
relation (see Table~\ref{tablefour}). Thus, the square of the 
effective mass in the medium relative to its free space value
(i.e., $M^{\star 2}/M^{2}$) should serve as an indicator of the
sensitivity of the approach to the various off-shell extrapolations.
Note that in a mean-field approximation to the Walecka model 
$M^{\star 2}/M^{2}$ is typically of the order of 40\%. Thus,
off-shell ambiguities are expected to be large
(see Table~\ref{tablefive}).

 The pseudovector contribution to the binding-energy difference 
of mirror nuclei is displayed, alongside the pseudoscalar 
results, in Table~\ref{tablefive}. As suggested previously, 
these results are very close to those obtained in 
Table~\ref{tablefour} using the free upper-to-lower ratio.
More significantly, however, is the very dramatic reduction 
in the binding-energy differences relative to the pseudoscalar 
results. Indeed, while a calculation with a pseudoscalar vertex 
accounts for about 70-85\% of the ONS anomaly, only 30\% of it can 
be explained with a pseudovector vertex. It is important to stress 
that although some theoretical guidelines do exist, off-shell 
ambiguities might be difficult to resolve given that the contribution 
from the isospin-violating part of the potential to hadronic 
observables is small.

\section{Conclusions}
\label{sec:concl} 

 Binding-energy differences of mirror nuclei have been computed 
in a model that includes two sources of charge-symmetry-breaking 
in the NN potential; isoscalar-isovector mixing in the meson 
propagator and isospin violation in the meson-nucleon coupling 
constants. For the vector-meson sector we have used a VMD-inspired 
model. Thus, on the basis of this assumption---but little else---we 
concluded that neither $\rho$-$\omega$ mixing nor the dominant 
vector-vector component of the vector-meson-exchange potential can 
contribute to the binding-energy discrepancy between theory and 
experiment. In models of this sort this conclusion seems 
unavoidable---as gauge invariance forces, both, the $\rho$-$\omega$ 
mixing amplitude and the isospin-violating vector coupling to vanish 
at $Q^{2}=0$. This is in contrast to previous analyses that employed 
the on-shell value of the $\rho$-$\omega$ mixing amplitude to explain 
the bulk of the ONS anomaly~\cite{miller90,bluiqb87,coon87,ishsas90}. 
Hence, in our model the anomaly must be explained exclusively in
terms of $\pi$-$\eta$ mixing and of isospin violation in the pion-, 
rho-, and omega-nucleon coupling constants.

 To compute the ${}^{3}$He-${}^{3}$H binding-energy difference a
nonrelativistic estimate of the ${}^{1}S_{0}$ component of the CSB 
potential was obtained. In our model all sources of CSB generate
a ${}^{1}S_{0}$ interaction with a long-range part that is 
attractive in the $pp$ channel and repulsive in the $nn$ channel. 
This result is robust, as it only depends on the sign of the 
$\pi$-$\eta$ mixing amplitude and on the sign of the 
isospin-violating meson-nucleon coupling constants; recall that 
in our model the latter is determined from the up-down quark mass 
difference, which we assume to be negative. The ${}^{3}$He-${}^{3}$H 
binding-energy difference, however, was seen to be very sensitive 
to the short-distance behavior of the potential. Indeed, the 
binding-energy difference emerged from a delicate interplay between 
the short- and long-range parts of the potential. This interplay 
introduces a node in the CSB potential which can be shifted to a 
large enough distance as to make the binding-energy difference 
almost consistent with experiment. We stress that the (almost)
agreement with experiment could only be achieved at the expense 
of introducing unrealistically soft form factors. 

 For the medium-mass (A=15--41) region we computed the binding-energy
difference of mirror nuclei in a relativistic mean-field approximation 
to the Walecka model. Due to the larger size of the system, relative 
to ${}^{3}$He, our results were fairly insensitive to the short-distance 
structure of the potential. Moreover, explicit spin-dependent effects 
are known to be small for these spin-saturated nuclei. As a consequence, 
the numerical impact from the vector-meson sector on the ONS anomaly was 
seen to be small. We proposed pseudoscalar-meson exchange as an 
alternate mechanism to explain the ONS anomaly in the medium-mass region. 
We computed binding-energy differences that were comparable---and in most 
cases larger---than those obtained from previous estimates using on-shell 
$\rho$-$\omega$ mixing. We have concluded that the pseudoscalar 
sector---alone---could explain about 70-85\% of the ONS anomaly. Two 
effects, ignored until now, were responsible for these findings: a) the 
inclusion of isospin violation in the pion-nucleon coupling constant, 
and b) relativity, which enhances the small components of the bound-state 
wave functions relative to their free-space value. We reached these 
conclusions by assuming a pseudoscalar coupling for, both, the $NN\pi$ 
and $NN\eta$ vertices. Yet other choices---all of them equivalent 
on-shell---are possible; such as pseudovector coupling. There is no 
definite prescription on how to take the CSB potential off-shell. 
Moreover, this issue is complicated further by the fact that chiral 
symmetry favors a pseudovector representation for the $NN\pi$ vertex 
while a recent analysis of $\eta$-photoproduction data seems to suggest 
a pseudoscalar $NN\eta$ coupling. Hence, to gauge the sensitivity of our 
results to the various off-shell extrapolations we repeated the calculation 
by adopting a pseudovector representation for both vertices. We concluded 
that the off-shell ambiguities were large; while 70-85\% of the anomaly 
could be explained with a pseudoscalar coupling only 30\% of it could
be accounted for with a pseudovector vertex.

 In conclusion, we have used a VMD-inspired model to examine the
effect of isospin-violating meson-nucleon coupling constants and
of $\pi$-$\eta$ mixing on the binding-energy difference of mirror
nuclei. We could account for most of the ${}^{3}$He-${}^{3}$H  
binding-energy difference---provided very soft form factors were
assumed. Moreover, we could explain the ONS anomaly in the 
medium-mass region---provided a pseudoscalar representation was 
adopted. Based on these findings we expect that the 
Okamoto-Nolen-Schiffer anomaly will remain---even after more 
than a quarter of a century---the source of considerable theoretical 
debate.

\acknowledgments
We thank S. Gardner, C.J. Horowitz, D. Robson, and B.D. Serot 
for many helpful conversations. We acknowledge the generosity
of E. Pace for providing us with the three-body wave function.
This work was supported by the DOE under Contracts Nos. 
DE-FC05-85ER250000, DE-FG05-92ER40750 and DE-FG05-86ER40273.

\begin{figure}
 \caption{The integrand for the ${}^{3}$He-${}^{3}$H 
          binding-energy difference arising from 
          one-pion exchange for three different choices
	  of form factors: $\Lambda_{NN\pi}\rightarrow\infty$
          (solid line), $\Lambda_{NN\pi}=1.7$~GeV (dashed line)
          $\Lambda_{NN\pi}=0.8$~GeV (dot-dashed line). The
	  one-pion exchange contribution to $\Delta E$, which 
          is the area under the appropriate curve, appears in 
          parentheses next to its label.}
 \label{figone}
\end{figure}

\begin{figure}
 \caption{The integrand for the ${}^{3}$He-${}^{3}$H 
          binding-energy difference arising from 
          one-omega exchange for three different choices
	  of form factors: $\Lambda_{NN\omega}\rightarrow\infty$
          (solid line), $\Lambda_{NN\omega}=1.85$~GeV (dashed line)
          $\Lambda_{NN\omega}=1.3$~GeV (dot-dashed line). The
	  one-omega exchange contribution to $\Delta E$, which 
          is the area under the appropriate curve, appears in 
          parentheses next to its label.}
 \label{figtwo}
\end{figure}

\mediumtext
 \begin{table}
  \caption{Meson masses, coupling constants, tensor-to-vector ratio,
           and cutoff parameters of the Bonn B potential.}
   \begin{tabular}{ccccc}
    Meson    &  Mass (MeV) & $g^{2}/4\pi$  & $f/g$ & $\Lambda$~(MeV) \\
     \tableline
    $\pi$    &     138     &     14.21     &  ---  &      1700       \\
    $\eta$   &     549     &      0.90     &  ---  &      1700       \\
    $\rho$   &     769     &      0.42     &  6.1  &      1850       \\
    $\omega$ &     783     &     11.13     &  0.0  &      1850       \\
   \end{tabular}
  \label{tableone}
 \end{table}

\mediumtext
 \begin{table}
  \caption{Contribution to the ${}^{3}$He-${}^{3}$H binding-energy 
           difference (in keV) arising from $\pi$-$\eta$ mixing, 
           $\pi-$, $\rho-$, and $\omega-$meson exchange for three 
           different values of the cutoff parameters; see text for 
           details. The experimental value is $\Delta E = 
           71\pm19\pm5$~keV. For comparison, we also include the 
	   contribution from on-shell $\rho$-$\omega$ mixing.}
   \begin{tabular}{cccc}
             & Point coupling & Hard form factors  & Soft form factors  \\
     \tableline
    $\pi$-$\eta$     &  $  +7.4$    &   $+13.6$    &  $+22.2$    \\
    $\pi$            &  $ -27.9$    &   $-16.2$    &  $+13.0$    \\
    $\rho$           &  $ -49.4$    &   $-37.3$    &  $-20.4$    \\
    $\omega$         &  $ -53.8$    &   $-40.9$    &  $-22.4$    \\
    Total            &  $-123.7$    &   $-80.8$    &  $ -7.6$    \\
     \tableline
    $\rho$-$\omega$  &  $ +57.9$    &   $+77.9$    &  $+87.3$    \\
   \end{tabular} 
  \label{tabletwo}
 \end{table}

\mediumtext
 \begin{table}
  \caption{Contribution to the binding-energy differences of mirror 
           nuclei (in keV) arising from isospin violation in the 
           pion-nucleon coupling constant, $\pi$-$\eta$ mixing, their 
           sum, and on-shell $\rho$-$\omega$ mixing. Also included are 
           the remaining differences ($\Delta$) between experiment and 
           the Coulomb energy computed in three different  
           models\protect{\cite{kkb96,sato76}}.}
   \begin{tabular}{cc|ccc|ccc|ccc}
       & & 
       \multicolumn{3}{c|}{Pseudoscalar}  &
       \multicolumn{3}{c|}{Rho-omega}     &  
       \multicolumn{3}{c} {ONS Anomaly}   \\ \hline
       A  &  $\phantom{xx}$State$\phantom{xx}$ &  
       $\delta g_{\scriptscriptstyle{NN}\pi}$  &
       $\pi$-$\eta$ & $\phantom{xx}$Total$\phantom{xx}$ & 
       Hartree & Fock & $\phantom{xx}$Total$\phantom{xx}$ 
       & $\phantom{x}$$\Delta_{\rm DME}$$\phantom{x}$
       & $\Delta_{\rm SKII}$$\phantom{x}$ 
       & $\Delta_{\rm REL}$$\phantom{x}$ \\ 
     \tableline
  15   &   $1p_{1/2}^{-1}$  &     201   &    145   &   346   &   379   
                            &  $-$149   &    230   &   380   &   290  
                            &      95  \\  
  17   &   $1d_{5/2}$       &      83   &     56   &   139   &   246   
                            &   $-$91   &    155   &   300   &   190 
                            &      79  \\ 
  27   &   $1d_{5/2}^{-1}$  &     223   &    148   &   371   &   445   
                            &  $-$176   &    269   &   480   &   490
                            &     551  \\ 
  29   &   $2s_{1/2}     $  &     165   &    108   &   273   &   305   
                            &   $-$91   &    214   &   290   &   240
                            &      34  \\ 
  31   &   $2s_{1/2}^{-1}$  &     202   &    134   &   336   &   355   
                            &  $-$119   &    236   &   540   &   560
                            &      39  \\ 
  33   &   $1d_{3/2}$       &     159   &    113   &   272   &   363   
                            &  $-$121   &    242   &   360   &   280
                            &     213  \\ 
  39   &   $1d_{3/2}^{-1}$  &     255   &    180   &   435   &   455   
                            &  $-$169   &    286   &   540   &   430
                            &     404  \\
  41   &   $1f_{7/2}     $  &     133   &     89   &   222   &   355   
                            &  $-$129   &    226   &   440   &   350
                            &     404  \\
   \end{tabular}
  \label{tablethree}
 \end{table}

\mediumtext
 \begin{table}
  \caption{Contribution to the binding-energy differences of mirror 
           nuclei (in keV) arising from isospin violation in the 
           pion-nucleon coupling constant, $\pi$-$\eta$ mixing, their 
           sum, and on-shell $\rho$-$\omega$ mixing. Included in 
           parenthesis are the corresponding quantities computed
           with lower components generated from the free-space 
           relation.}
   \begin{tabular}{cccccc}
   A  &  State  &  $\delta g_{\scriptscriptstyle{NN}\pi}$ 
      & $\pi$-$\eta$ & Total & Rho-omega                    \\
     \tableline
  39   &   $1d_{5/2}^{-1}$  &   205~(103)   &   142~(73)   
                            &   347~(176)   &   294~(291)    \\
       &   $2s_{1/2}^{-1}$  &   233~(84)    &   158~(58)   
                            &   391~(142)   &   283~(271)    \\
       &   $1d_{3/2}^{-1}$  &   255~(92)    &   180~(65)   
                            &   435~(157)   &   286~(275)    \\
  41   &   $1f_{7/2}     $  &   133~(78)    &    89~(53)   
                            &   222~(131)   &   226~(224)    \\ 
   \end{tabular}
  \label{tablefour}
 \end{table}

\mediumtext
 \begin{table}
  \caption{Contribution to the binding-energy differences of mirror 
           nuclei (in keV) arising from isospin violation in the 
           pion-nucleon coupling constant, $\pi$-$\eta$ mixing, and
           their sum. The first(second) set of numbers were computed
           using a pseudoscalar(pseudovector) $NN\pi$ vertex.}
   \begin{tabular}{cc|ccc|ccc}
       & & 
       \multicolumn{3}{c|}{Pseudoscalar}  &
       \multicolumn{3}{c}{Pseudovector}  \\ \hline
       A  &  $\phantom{xx}$State$\phantom{xx}$ &  
       $\delta g_{\scriptscriptstyle{NN}\pi}$  &
       $\pi$-$\eta$ & $\phantom{xx}$Total$\phantom{xx}$ &
       $\delta g_{\scriptscriptstyle{NN}\pi}$  &
       $\pi$-$\eta$ & $\phantom{xx}$Total$\phantom{xx}$ \\
     \tableline
  15   &   $1p_{1/2}^{-1}$  &     201   &    145   &   346   
                            &      67   &     49   &   116   \\
  17   &   $1d_{5/2}$       &      83   &     56   &   139   
                            &      52   &     36   &    88   \\
  27   &   $1d_{5/2}^{-1}$  &     223   &    148   &   371   
                            &     109   &     74   &   183   \\
  29   &   $2s_{1/2}     $  &     165   &    108   &   273   
                            &      54   &     36   &    90   \\  
  31   &   $2s_{1/2}^{-1}$  &     202   &    134   &   336   
                            &      62   &     43   &   105   \\ 
  33   &   $1d_{3/2}$       &     159   &    113   &   272   
                            &      49   &     35   &    84   \\
  39   &   $1d_{3/2}^{-1}$  &     255   &    180   &   435   
                            &      83   &     59   &   142   \\
  41   &   $1f_{7/2}     $  &     133   &     89   &   222   
                            &      74   &     51   &   125   \\
   \end{tabular}
  \label{tablefive}
 \end{table}

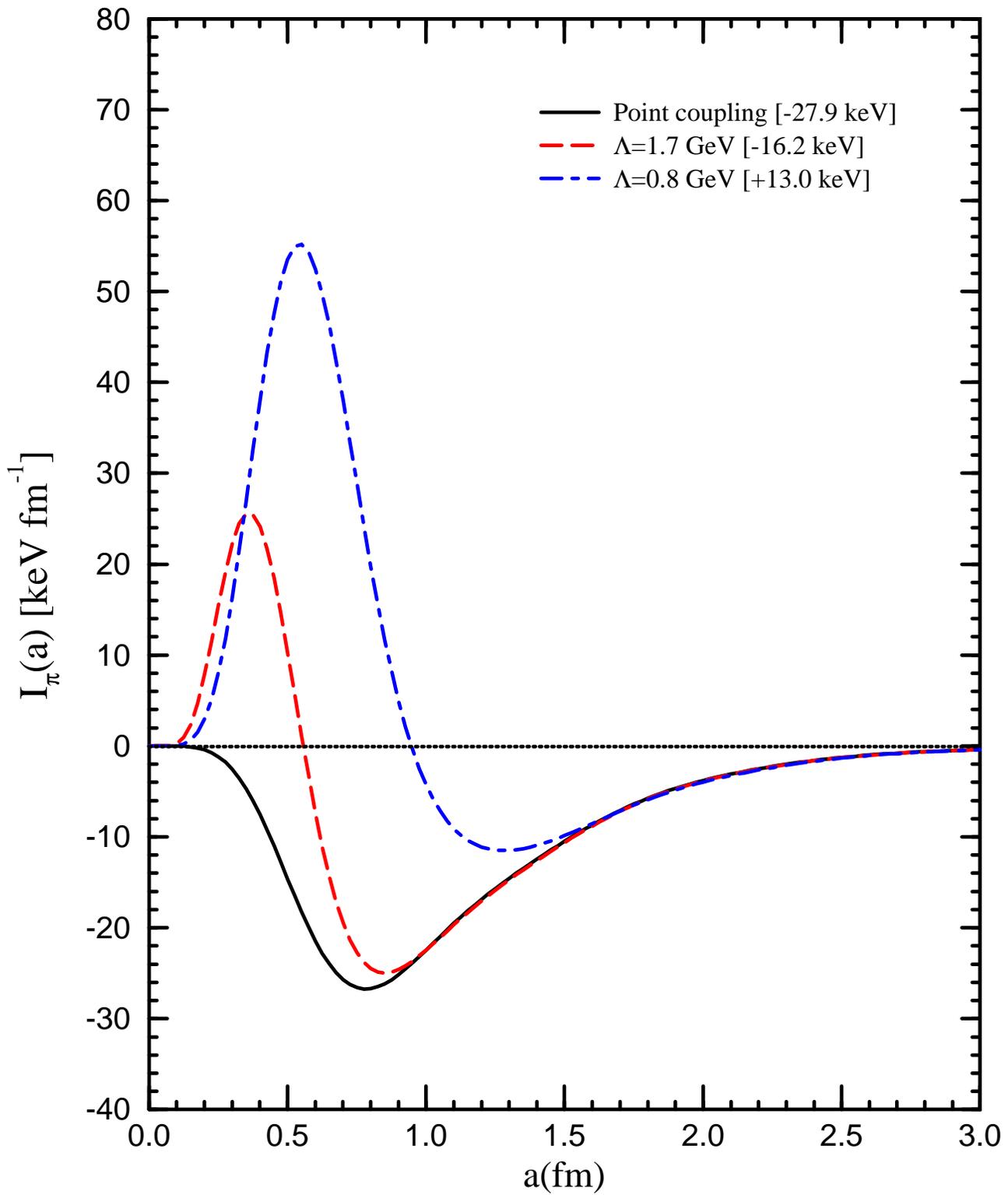
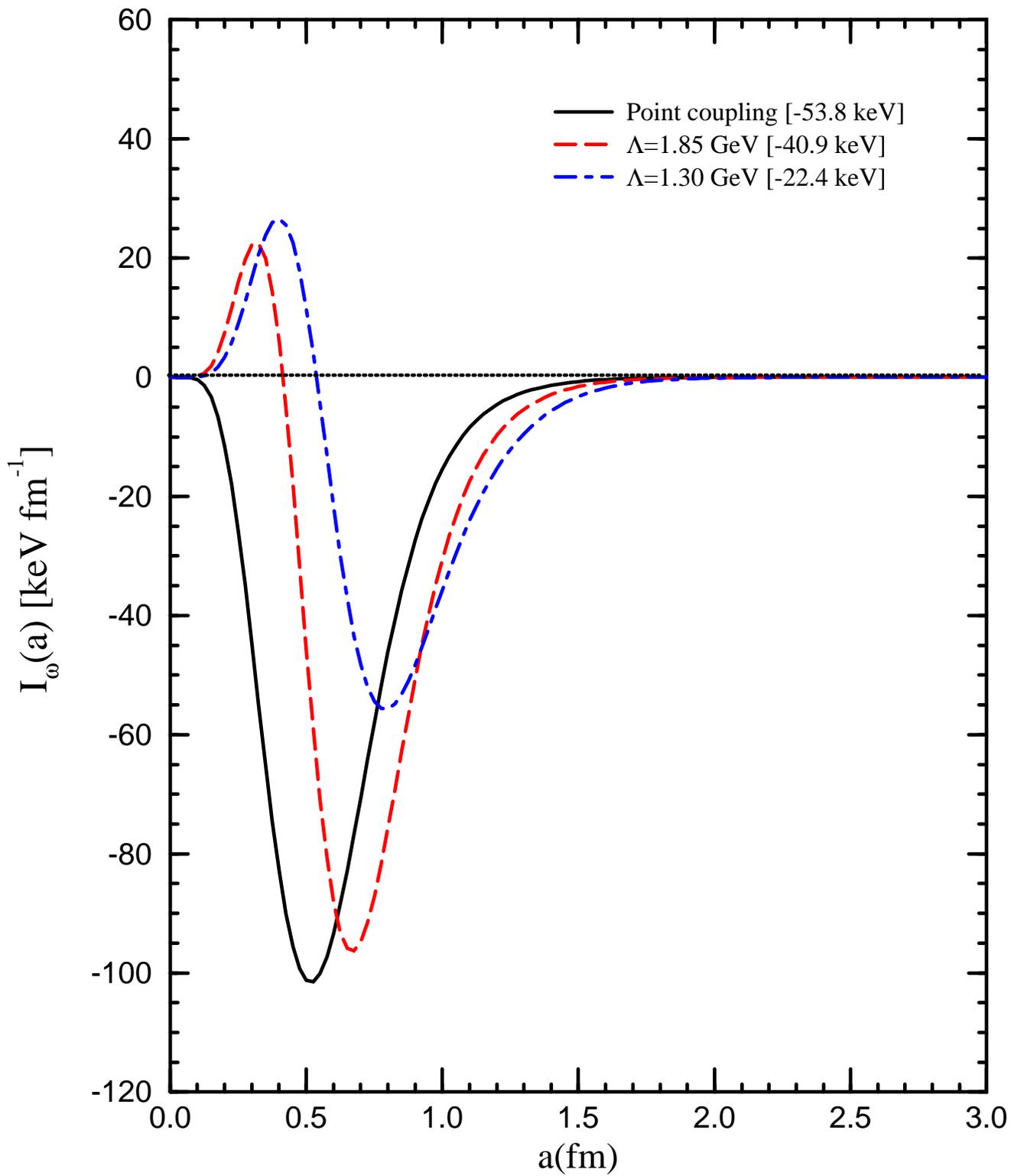
\end{document}